\documentclass[12pt]{article}
\usepackage{amssymb}
\usepackage[mathscr]{eucal}
\usepackage{url}
\usepackage{a4}
\usepackage{cite}
\usepackage{graphicx}
\usepackage{psfrag}
\usepackage{subfigure}
\usepackage{latexsym}
\usepackage{amsmath}
\newcounter{mnotecount}[section]

\begin{document}
\newcommand{\g}{$\bf \bar g g^{\alpha\beta}$}
\title{Relativistic Elasticity II}

 \author{Robert Beig\\ Faculty of Physics, University of  Vienna, Austria\\}

\maketitle
\begin{center}
\it{Dedicated to the memory of Bernd G. Schmidt}
\end{center}
\begin{abstract}
This paper is based on a series of talks given at the ESI program on 'Mathematical Perspectives of Gravitation Beyond the Vacuum Regime' in February 2022. It is meant to be an introduction to the field of relativistic elasticity for
readers with a good base in the mathematics of General Relativity with no necessary previous of knowledge of elasticity either in the classical or relativistic domain.
Despite its introductory purpose, the present work has new material, in particular related to the formal structure of the theory.
\\


\end{abstract}
\section{Introduction}
Elasticity theory is the branch of continuum mechanics dealing with deformable solids, which are ideal in the sense that dissipative processes can be ignored.
Good introductions to the nonrelativistic theory for the mathematically oriented reader are Ciarlet \cite{ciarlet2021mathematical} and Marsden, Hughes \cite{marsden1994mathematical}. \\
In the context of relativity elasticity was first treated (in special relativity) by Herglotz \cite{herglotz1911mechanik} as early as 1911. For many years, with papers by many different authors, the subject remained somewhat diffuse until the insightful work of Carter and Quintana \cite{carter1972foundations}, which became particularly influential among authors seeking astrophysical applications \cite{karlovini2003elastic}. The paper \cite{carter1972foundations} had been remarkably silent about what should be considered the basic dynamical variables of the theory. Then, in the work of Kijowski and Magli \cite{kijowski1992relativistic}, came the realization that the relativistic version of the classical theory of hyperelastic materials is fundamentally a theory - derivable from an action principle - of maps subject to some conditions, which are sometimes called back-to-label maps and called here configurations, from spacetime $M$ into a fictitious 3-dimensional space called material space\footnote{A lucid, but largely unnoticed, account had previously been given in the textbook of Soper \cite{soper1976classical}.}. The recent work of Brown \cite{brown2021elasticity}, similarly to Herglotz, describes the theory in the material (often called 'Lagrangian') perspective with configurations replaced by time dependent maps from material space into physical space relative to some slicing of spacetime.\\
The present paper should be viewed as an introduction to the geometric structure of the theory. A guiding principle throughout is the role of invariance under spacetime diffeomorphisms. This gives rise to a considerable departure from practically all the standard treatments of nonrelativistic elasticity. So, for example, by working with a diffeomorphism invariant theory on a general spacetime, the issue of 'material frame indifference' (see \cite{ciarlet2021mathematical} and \cite{marsden1994mathematical}) is completely bypassed. Furthermore the standard literature sees the theory primarily as a branch of mechanics, which results in a preference of the material picture.\\
In contrast we view elasticity as a field theory on spacetime, which makes the spacetime ('Eulerian') - as opposed to the material - viewpoint, where diffeomorphism invariance is manifest, the method of choice. Moreover: for whatever field is present in the vacuum region surrounding the body, the material picture is a priori not even defined. On the other hand the natural (in a sense described later) boundary conditions at the matter-vacuum interface render this a free boundary, and this, for the actual solution of problems, makes an intermediate use of the material picture - in which this boundary is fixed - indispensable.\\
Our treatment will be formal in the sense that there will be no function spaces - and also in the sense of staying away from astrophysical applications (for a recent account see e.g. \cite{penner2011tidal}) or detailed equations of state suitable for describing compact elastic bodies in astrophysics.\\
The plan of this paper is as follows. In Sect.2 we give a quick account of the Lagrangian theory of maps from spacetime to some other manifold. This has applications ranging from wave maps to continuum mechanics. In Sect.3 we study the kinematics of relativistic elasticity, namely the geometry of configurations. In Sect.4 we apply the results of Sect.2 to a theory of configurations, namely relativistic elasticity. Barotropic perfect fluids turn out to be a special case. But fluids in the Eulerian picture are usually described by the 4-velocity, energy density and pressure, all viewed as fields on $M$, which obey a 1st order system of partial differential equations (PDE's). There is no material manifold and no configurations.\\
 In Sect.5 we describe a formulation in this spirit, due to Tahvildar-Zadeh \cite{Tahvildar1998relativistic}, of elasticity in the so-called isotropic case. The next two sections concern the subjects of the nonrelativistic limit and linearization at a stressfree state. In Sect.8 there is a quick account of causality and local wellposedness, which largely draws on \cite{christodoulou2000action}. In Sect.9 we treat conditions at the matter-vacuum interface appropriate for freely floating bodies. Our treatment is based on the variational identity (\ref{basic2}), which is in turn based on (\ref{remarkable}). These, to the best of our knowledge, are new. The resulting boundary condition takes the form of the vanishing of normal stress on an a priori unknown timelike hypersurface in spacetime, namely the boundary of the world tube swept out by the congruence of timelike curves of the particles making up the material manifold. In Sect.10 we for completeness derive the elastic action in the material picture, although that is not used explicitly in the present work. In Sect.11 we review some time independent problems which have so far been solved. Finally, in Sect.12,
we give a short description of what is known  so far about the fully coupled Einstein-elastic system. These last two sections exclusively treat the case of finite bodies surrounded by vacuum.\\
Note, finally, that we have tried our presentation to be mostly self-contained - with the exception of sections 8,11 and 12 which have review character.
\section{Lagrangian theory of maps}
A careful introduction to the Lagrangian theory of maps is given in \cite{christodoulou2000action}, see also \cite{wong2011regular}. Absent from these works is however a discussion of the role of diffeomorphism invariance, which is a main focus here. The material in the present section is largely a generalization of \cite{beig2003relativistic}.\\
Let $f^A (x^\mu) \,,A= 1,..m$ be maps from spacetime $(M,g_{\mu\nu})$ with $g$ of signature $(-+++)$ to an $m$-dimensional manifold $N$.
The Lagrange function $L$ should be $L = L (f^A, \partial_\mu f^A; g^{\mu\nu})$. For the time being think of $N$ as $\mathbb{R}^m$ or a manifold in a fixed chart.
Our basic requirement is: $L$ is a scalar under general diffeomorphisms of $M$. Infinitesimally this means that
\begin{equation}\label{inf}
\xi^\mu \partial_\mu\, L = \frac{\partial L}{\partial f^A} \mathcal{L}_\xi f^A (x) + \frac{\partial L}{\partial \partial_\mu f^A} \mathcal{L}_\xi \partial_\mu f^A (x) + \frac{\partial L}{\partial g^{\mu\nu}}\mathcal{L}_\xi g^{\mu\nu}
\end{equation}
Expanding the left side in (\ref{inf}) and using the arbitrariness of the vector field $\xi$, we find that
\begin{equation}\label{fund}
(\partial_\mu f^A) \frac{\partial L}{\partial_\nu f^A} = 2 g^{\lambda \nu}\frac{\partial L}{\partial g^{\lambda\mu}}
\end{equation}
The relation (\ref{fund}) is in particular satisfied when
\begin{equation}
L = L(f^A, H^{BC})\hspace{0.5cm}\mathrm{where}\hspace{0.4cm}H^{AB} = (\partial_\mu f^A)(\partial_\nu f^B) g^{\mu\nu}
\end{equation}
Suppose $H^{AB}$ is non-degenerate, which in particular entails that $f$ is a submersion, i.e. $\partial_\mu f^A$ has maximal rank (in particular $m \leq 4$). Then (\ref{fund}) actually implies $L = L(f^A, H^{BC})$. To see this first note that in that case the tangent space of $(M,g)$  at the point $x$ in the fiber over $X = f(x)$ splits into an $(n - m)$ dimensional vertical (i.e. spanned by vectors in the null space of $\partial_\mu f^A$) distribution  and an orthogonal $m$ dimensional horizontal distribution, and $H^{AB}$ can be seen as the inverse metric on the horizontal distribution. But (\ref{fund}) implies that
\begin{equation}\label{mixed}
\frac{\partial L}{\partial g^{\mu\nu}}v^\nu = 0
\end{equation}
for $v^\mu$ any vertical direction. Thus $L$ can depend on $g^{\mu\nu}$ only via $H^{AB}$, so $L = L(f,^A, \partial_\mu f^B, H^{CD})$. Inserting this back into (\ref{fund}) yields that actually $L = L(f^A, H^{BC})$, which ends the proof.\\
Note that (\ref{mixed}) also means that $T_{\mu\nu}$ has no mixed vertical-horizontal components. The identity (\ref{fund}) has further important implications.\\
{\bf{Theorem}:}\\
{\bf{(F1)}} the canonical stress-energy tensor $\mathcal{T}_\mu{}^\nu = (\partial_\mu f^A) \frac{\partial L}{\partial_\nu f^A} - L \,\delta_\mu{}^\nu$ satisfies $g_{\nu\rho} \mathcal{T}_\mu{}^\rho = T_{\mu\nu}$, where $T_{\mu\nu} = 2 \frac{\partial L}{\partial g^{\mu\nu}} - L g_{\mu\nu}$, that is the r.h. side of the Einstein equations.
This identity is a special case of the Belinfante-Rosenfeld theorem (see \cite{kijowski1997unconstrained} and references therein.)\\
{\bf{(F2)}} The Euler-Lagrange equations for the action $S = \int_M L \sqrt{- g}\, d^4 x$ imply
$\nabla_\nu T_\mu{}^\nu = 0$ and the latter 4 equations are in general not independent.
More precisely, there holds the (Noether-type) 'off-shell' identity (see also \cite{kijowski1997unconstrained})
\begin{equation}\label{Bianchi}
\nabla_\nu T_\mu{}^\nu = - (\partial_\mu f^A) \, \mathcal{E}_A\,,
\end{equation}
where
\begin{equation}
- \mathcal{E}_A = \frac{1}{\sqrt{- g}}\,\partial_\mu \left(\sqrt{- g} \, \frac{\partial L}{\partial_\mu f^A}\right) - \frac{\partial L}{\partial f^A}
\end{equation}
In the case $f$ is a submersion, i.e. $\partial_\mu f^A$ has rank $m = \mathrm{dim} N$, $\nabla_\nu T_\mu{}^\nu = 0$ is actually sufficient for $\mathcal{E}_A = 0$.\\
{\bf{Proof}:} {\bf{(F1)}} is a straightforward verification based on (\ref{fund}). {\bf{(F2)}} is also easy, using $\mathcal{T}_\mu{}^\nu$ on the left side.\\
One way to think of the identity (\ref{Bianchi}) is that it is a direct expression of diffeomorphism invariance of the action in the same way as the contracted Bianchi identity for the Einstein tensor is a consequence of the diffeomorphism invariance of the Hilbert action. To see this
vary the action $S[f,\partial f;g]$ under a 1-parameter family of diffeomorphisms generated by a vector field $\xi^\mu$, there results
\begin{equation}\label{diffeo}
0 = \frac{1}{2}\int T_{\mu\nu} (\mathcal{L}_\xi g^{\mu\nu}) \sqrt{- g} \,d^4x + \frac{d}{d \epsilon}\vert_{\epsilon = 0}\, S[f + \epsilon \mathcal{L}_\xi f;g]\,,
\end{equation}
where integration is carried out over some fixed domain in spacetime and $\xi^\mu$ is compactly supported there. Now the 2nd term in (\ref{diffeo}) equals $\int \mathcal{E}_A (\xi^\mu \partial_\mu f^A) dV_g$. Using that $\mathcal{L}_\xi g^{\mu\nu} = - 2 \nabla^{(\mu} \xi^{\nu)}$ and integrating by parts the 1st term in (\ref{diffeo}) gives $\int (\nabla^\nu T_{\mu\nu}) \xi^\mu dV_g$. So (\ref{Bianchi}) follows using the arbitrariness of $\xi^\mu$.\\
{\bf{(F2)}} implies: Provided that the  matter system has a well-posed Cauchy problem in the background of any given spacetime ('test case'), the combined matter-Einstein system is also well-posed (see e.g.\cite{ringstrom2009cauchy}).\\

Clearly
\begin{equation}\nonumber
- \mathcal{E}_A = \frac{\partial^2 L}{\partial (\partial_\mu f^A)\partial (\partial_\nu f^B)} \,\partial_\mu \partial_\nu f^B + \mathrm{l.o.}
\end{equation}
More explicitly (setting $M^{\mu\nu}_{AB} = \frac{\partial^2 L}{\partial (\partial_\mu f^A)\partial (\partial_\nu f^B)}$)
\begin{equation}\nonumber
M^{\mu\nu}_{AB} = 2\, \frac{\partial L}{\partial H^{AB}}\,g^{\mu\nu} + 4\,\frac{\partial^2 L}{\partial H^{AC}\partial H^{BD}}(\partial_\rho f^C)(\partial_\sigma f^D)g^{\mu\rho}g^{\nu\sigma}
\end{equation}
Note: only $M^{(\mu\nu)}_{\,AB}$ contributes to the equations of motion (EOM).\\
So far the $f^A$'s were maps to some manifold $N$ in a fixed coordinate system $X^A$: $f^A: x^\mu \in M \mapsto X^A = f^A(x^\mu) \in N$.
We consider 2 important cases:\\
(i) Homogenous case: $N$ is viewed as an affine vector space and $f^A$ a collection of scalars subject only to affine transformations. Suppose furthermore that $L = L(H^{AB})$. Then the exact Euler-Lagrange expression can be written as
\begin{equation}\nonumber
- \mathcal{E}_A = M^{\mu\nu}_{AB}\nabla_\mu \partial_\nu f^B
\end{equation}

Proof: note $M^{\mu\nu}_{AB} (f, \partial f; g)$ transforms as a $(2,0)$-tensor under spacetime coordinate transformations and $\mathcal{E}_A (f,\partial f, \partial^2 f; g, \partial g)(x)$ as a scalar. Then, taking normal coordinates centered at $x$ gives the result. The 'fully explicit' proof is remarkably tricky.\\

(ii) Isotropic case: the target space $N$ is a manifold equipped with a Riemannian metric $G_{AB}(X)$, and $L$ is a function of the principal invariants (or what is the same: eigenvalues) of $H^A{}_B$, where
$H^{D}{}_A = G_{AC}(f)(\partial_\mu f^C) (\partial_\nu f^D) g^{\mu\nu}$. For example, when $m=3$, the Lagrangian is then of the form
\begin{equation}
L = (H^A{}_A, H^A{}_B H^B{}_A, \mathrm{det}(H^A{}_B))
\end{equation}
Then the theory is also invariant under diffeomorphisms of $N$. Furthermore the exact Euler-Lagrange expression is
\begin{equation}
- \mathcal{E}_A = M^{\mu\nu}_{AB}\overline{\nabla}_\mu \partial_\nu f^B
\end{equation}
where\\
\vspace{0.2cm}
\begin{equation}
\overline{\nabla}_\mu \partial_\nu f^A = \partial_\mu \partial_\nu f^A - \Gamma^\rho_{\mu\nu} \partial_\rho f^A + \Gamma^A_{\!BC} (\partial_\mu f^B)(\partial_\nu f^C)\,,
\end{equation}
where $\Gamma^A_{\!BC}$ are the Christoffel symbols of $G_{AB}$. The isotropic and homogenous case is where $G_{AB} = \delta_{AB}$. Suppose, in the isotropic case, we take $L = H^A{}_A = G_{AB}(f)H^{AB}$. The EOM is then linear in $f$ when $G_{AB}$ is flat and semilinear otherwise - these are called {\it{wave maps}} in the mathematics \cite{tataru2004wave} and {\it{sigma models}} in the physics \cite{ketov2009nonlinear} literature.
One can, as done here, also consider the general isotropic case. This happens e.g. for the Skyrme model without the mass term \cite{skyrme1994selected} or isotropic relativistic elasticity. In this case the EOM are quasilinear.
\section{Geometry of configurations}
We now specialize to the case $N = \mathcal{B}$ called material manifold or body, a domain in $\mathbb{R}^3$ with smooth boundary or all of $\mathbb{R}^3$, thought of as the collection of elements making up the material. Note: no further structure is imposed on $\mathcal{B}$ to start with. For bodies of everyday life that is defined by the 'making' of the body, for astrophysical objects it is largely determined by gravity.\\
{\bf{Definition:}} The field $f: (M,g) \rightarrow \mathcal{B}$ is called a configuration if $\partial_\mu f^A$ has everywhere maximal rank and $f$ is surjective with level sets timelike curves in $M$.\\
 In other words: a configuration is a surjective submersion
 with timelike fibers. Codimension 1 means we are dealing with media composed of pointlike objects ('particles'). Timelike fibers means we are dealing with massive particles. \\
Given a configuration $f^A$, there is then - subject to time orientation - a unique vector field $u^\mu$, s.th.
 \begin{equation}
 u^\mu \partial_\mu f^A = 0\,,\hspace{1cm}g_{\mu\nu} u^\mu u^\nu = - 1\,.
 \end{equation}

 Note that the concept of configuration is intrinsically nonlinear.\\
 Let $h_{\mu\nu} = g_{\mu\nu} + u_\mu u_\nu$ be the metric on the 'horizontal distribution' defined by $u^\mu$. There then exists a unique field $\psi^\mu{}_A$ with $\psi^\mu{}_A u_\mu = 0$ such that
 \begin{equation}
\psi^\mu{}_A \,\partial_\nu f^A = h^\mu{}_\nu\,,\hspace{1cm}\psi^\mu{}_B \,\partial_\mu f^A = \delta^A{}_B\,.
\end{equation}
Furthermore $H^{AB} = (\partial_\mu f^A)(\partial_\nu f^B) g^{\mu\nu}$ is now positive definite.\\
Note: $H^{AB}, u^\mu, h_{\mu\nu}, \psi^\mu{}_A$ are all algebraic functions of $(\partial f,g)$.
Some representative formulae in this regard are:

\begin{equation}\label{h}
\psi^\mu{}_A \psi^\nu{}_B\, H^{AB} = h^{\mu\nu}\,,\hspace{0.5cm}\partial_\nu f^A = \psi^\mu{}_B\, h_{\mu\nu}\, H^{AB}
\end{equation}

\begin{equation}
\frac{\partial u^\mu}{\partial (\partial_\nu f^A)} = - u^\nu \psi^\mu{}_A\,,\hspace{1.4cm}\frac{\partial \psi^\mu{}_A}{\partial (\partial_\nu f^B)} = - \psi^\mu{}_B \psi^\nu{}_A - H_{AB} u^\mu u^\nu\,,
\end{equation}
where $H_{AB} H^{BC} = \delta^C{}_A$ \,or
\begin{equation}
\frac{\partial u^\mu}{\partial g^{\nu\rho}} = - \frac{1}{2}u^\mu u_\nu u_\rho\,,\hspace{1.5cm}\frac{\partial \psi^\mu{}_A}{\partial g^{\rho\sigma}}=
- u^\mu h_{\nu(\rho} u_{\sigma)}\psi^\nu{}_A
\end{equation}
We will later write down some coordinate formulae.
\subsection{Uncontorted configurations}
In field theory time independent fields are often used as reference fields. For configurations this means that there is a timelike  Killing field $\xi^\mu$ on $(M,g)$ such that $\mathcal{L}_\xi f^A = \xi^\mu \partial_\mu f^A = 0$, which in turn says that the particle flow defined by $f^A$ is tangent to Killing trajectories: $u^\mu \sim \xi^\mu$ with $\xi$ a Killing field on $(M,g)$. In the case of configurations there is available a slightly weaker concept, which, for reasons explained below, we call uncontorted configurations and which is the analogue of Riemannian submersions in the case $g_{\mu\nu}$ is Riemannian. Uncontorted configurations are as close as configurations can get to being isometries.\\
{\bf{Definition:}} A configuration is called uncontorted if $u^\mu$ is Born-rigid (see \cite{Pirani1962rigid}), i.e.
\begin{equation}
\mathcal{L}_u h_{\mu\nu} = 0
\end{equation}
Clearly time independence implies uncontortedness. To explain the relation with more standard terminology recall the decomposition
\begin{equation}
h_\mu{}^{\mu'} h_\nu{}^{\nu'}\, \nabla_{(\mu'} u_{\nu')} = \sigma_{\mu\nu} + \frac{1}{3} h_{\mu\nu} \Theta
\end{equation}
into shear $\sigma_{\mu\nu}$ and divergence $\Theta = \nabla_\mu u^\mu$. So a configuration is uncontorted iff its congruence of particle world lines is shear- and expansionfree.
To explain the geometric meaning of non-contortedness calculate
\begin{equation}
\mathcal{L}_u h_{\mu\nu} = \mathcal{L}_u [(\partial_\mu f^A)(\partial_\nu f^B H_{AB})] = (\partial_\mu f^A)(\partial_\nu f^B)
 u^\rho \partial_\rho H_{AB}\,,
\end{equation}
where we were using (\ref{h}) in the first equality and $[\mathcal{L},d] = 0$ in the second. So uncontortedness is equivalent to

\begin{equation}
u^\mu \partial_\mu H_{AB} = 0\hspace{0.5cm}\mathrm{or}\hspace{0.5cm}u^\mu \partial_\mu H^{AB}= 0\,,
\end{equation}
which in turn means that there exists a Riemannian metric $G_{AB}(X)$ on $\mathcal{B}$ so that $H_{AB}(x)= G_{AB}(f(x))$, i.e. $h_{\mu\nu}$ is the pull-back of G under $f$. \\
In a similar fashion a configuration whose particle flow is shearfree defines a conformal metric on $\mathcal{B}$ which pulls back to the conformal metric defined by $h$, and a configuration which is incompressible (i.e. has non-expanding flow: $\Theta = 0$) defines a volume form $\Omega_{ABC}$ on $\mathcal{B}$ which under $f$ pulls back to the 3-form $\epsilon_{\mu\nu\rho\sigma} u^\sigma$ on $M$. Shear and expansion are pieces of information referring to the horizontal distribution defined by $f$. The remaining pieces of horizontal data are the acceleration $u^\nu \nabla_\nu u^\mu$ and  rotation $\omega_{\mu\nu}$ defined by
\begin{equation}
\omega_{\mu\nu} = h_\mu{}^{\mu'} h_\nu{}^{\nu'}\, \nabla_{[\mu'} u_{\nu']}\,,
\end{equation}
which is the obstruction to integrability of the horizontal distribution.\\
Note finally that non-contorted configurations which are not time independent are rare, since Born-rigid congruences are generically Killing. For example it is a theorem of Herglotz and Fritz Noether (see \cite{Pirani1962rigid}) that a congruence on Minkowski space which is Born-rigid and with non-vanishing rotation has to be Killing.
\subsection{Back and forth between $\mathcal{B}$ and $M$ under a configuration}
The relation between $h_{\mu\nu}$ and $G_{AB}$ in the previous subsection is a special case of the following observation: Let $\Omega_{A..}$ be a covariant tensor field on $\mathcal{B}$. Then its pull-back $(f^\star \Omega)_{\mu..}$ under a configuration $f^A$ is a covariant tensor field on $M$ of the same type and is horizontal, i.e. gives zero under contraction with $u^\mu$ over any index. Furthermore $(\mathcal{L}_u f^\star \Omega)_{\mu..} = 0$. The latter fact can be easily checked explicitly by noting that
$\mathcal{L}_u \partial_\mu f^A = 0$.\\
Conversely, let $o_{\mu..}$ be a covariant, horizontal tensor field on $M$ with $\mathcal{L}_u o_{\mu..}$ vanishing for the flow vector field $u^\mu$ corresponding to some configuration $f^A$. Then $o_{\mu..}$ arises as pull back of a covariant tensor field $\Omega_{A..}$. To see this pick a hypersurface $\Sigma$ in $M$ which is everywhere transversal to $u^\mu$, e.g. spacelike. Thus the map $\bar{f}_\Sigma: \Sigma \rightarrow \mathcal{B}$ is invertible and onto. Define a tensor field $\Omega_{A..}$ on $\mathcal{B}$ by pull back of $o_{\mu..}|_\Sigma$ under $\bar{f}^{- 1}$. The pull back of $\Omega_{A..}$ along all of $M$ is Lie derived under $u^\mu$ and coincides with $o_{\mu..}$ initially, thus everywhere. This ends the proof.

\subsection{Some coordinate identities}
Note first the map $f^A$ is a configuration onto its image, iff $H^{AB}$ is positive definite and, in local coordinates $(t, x^i)$ with $t=$ constant being spacelike hypersurfaces, the matrix $\partial_i f^A$ is nonsingular\footnote{Actually it is not hard to see that the latter condition follows from the former.}.
\\
The coordinate velocity $v^\mu \partial_\mu = \partial_t + V^i \partial_i$, since
$v^\mu \partial_\mu f^A = 0$, is in terms of the configuration given by
$V^i = - \phi^i{}_A \,\partial_t f^A$, where $\phi^i{}_A\, \partial_i f^B = \delta^B{}_A$. \\
Take more specifically ADM coordinates, i.e.$(t, x^i)$, where $g(\partial_t - Y^j \partial_j, \partial_i) = 0$. Setting $g^{- 1}(dt, dt) = - N^{-2}$ and $g(\partial_i, \partial_j) = g_{ij}$:
\begin{equation}
g_{\mu\nu} dx^\mu dx^\nu = - N^2 dt^2 + g_{ij}(dx^i + Y^i dt)(dx^j + Y^j dt)
\end{equation}
\begin{equation}
g^{\mu\nu}\partial_\mu \partial_\nu = - \frac{1}{N^2} (\partial_t - Y^i \partial_i)^2 + g^{ij}\partial_i \partial_j
\end{equation}
\begin{equation}
u^\mu \partial_\mu = \frac{(\partial_t - Y^i \partial_i) + W^i \partial_i}{(N^2 - W^2)^\frac{1}{2}}\,,\hspace{1cm}W^i =  Y^i + V^i
\end{equation}
and $W^2 = W^i W^j g_{ij} = W^i W_i$. We call the  quantity $W^i$ relative shift, following a suggestion  by L.Andersson (private communication). Furthermore\\
\begin{equation}
(\partial_\mu f^A) dx^\mu = (\partial_i f^A)[- W^i dt + (dx^i + Y^i dt)]
\end{equation}

\begin{equation}
H^{AB} = (\partial_i f^A)(\partial_j f^B)\, \left(g^{ij} - \frac{W^iW^j}{N^2}\right)
\end{equation}

\begin{equation}
u_\mu dx^\mu = \frac{- N^2 dt + W_i(dx^i + Y^i dt)}{(N^2 - W^2)^\frac{1}{2}}
\end{equation}

\begin{equation}
\psi^\mu{}_A\, \partial_\mu = \phi^i{}_A \left[\frac{W_i}{N^2 - W^2} (\partial_t - Y^l\partial_l) + \left(\delta^j{}_i + \frac{W^j W_i}{N^2 - W^2}\right)\partial_j\right]
\end{equation}
\subsection{Galilean configurations}
A Galilean spacetime is affine $\mathbb{R}^4$ with
the degenerate contravariant metric $e^{\mu\nu} \partial_\mu \partial_\nu = \delta^{ij} \partial_i \partial_j$ and covector $\tau_\mu dx^\mu = dt$ annihilated by $e^{\mu\nu}$. Affine transformations leaving $M = (\mathbb{R}^4, e,\tau)$ invariant form the 10 dimensional Galilei group.
\\
Associated with a configuration $f: M \rightarrow \mathcal{B}$ is the 4-velocity $v^\mu \partial_\mu$, uniquely given by $v^\mu \partial_\mu f^A = 0$ and $v^\mu \tau_\mu = 1$, so
\begin{equation}\label{recall}
v^\mu \partial_\mu = \partial_t + V^i \partial_i = \partial_t - \phi^i{}_A \, \partial_i f^A\,,\,\,\mathrm{where}\,\,\,\phi^i{}_A \,\partial_j f^A = \delta^i{}_j
\end{equation}
and
\begin{equation}
(\partial_\mu f^A) dx^\mu = (\partial_i f^A)(dx^i - V^i dt)
\end{equation}
The field $\psi^\mu{}_A$ is now defined by
\begin{equation}
\psi^\mu{}_A (\partial_\nu f^A) = \delta^\mu{}_\nu - v^\mu \tau_\nu
\end{equation}
which gives
\begin{equation}
\psi^\mu{}_A \partial_\mu = \phi^i{}_A \partial_i
\end{equation}
Note: both $\psi^\mu{}_A$ and $K^{AB} = (\partial_\mu f^A)(\partial_\nu f^B)e^{\mu\nu}$ depend only on spatial derivatives of $f$.
\\
The analogue of uncontorted configurations are now those for which $v^\mu$ leaves $e^{\mu\nu}$ and $\tau_\mu$ invariant. These are the infinite dimensional group of rigid body motions, i.e. a time dependent rotations plus time dependent spatial translations plus time translations. Like in the relativistic case, the uncontorted configurations considered here will always be ones constant along time translations or rigid rotations.
\subsection{Time independent configurations}
For a time independent background the ADM variables $(N, Y^i, g_{jk})$ are unsuitable. We rather take $(V, Y_i, h_{jk})$ defined by
\begin{equation}
V^2 = - \xi^\mu \xi^\nu g_{\mu\nu} = N^2 - Y^2,\,\,\,Y_i = g_{ij} Y^j,\,\,\, h_{ij} = g_{ij} - \frac{1}{N^2} Y_i Y_j
\end{equation}
whence
\begin{equation}\label{geroch}
g_{\mu\nu} dx^\mu dx^\nu = - V^2(dt + Y_i dx^i)^2 + h_{ij} dx^i dx^j
\end{equation}
Geometrically $h_{ij}$ is the metric on the space $Q$, which is the quotient space of $(M,g_{\mu\nu})$ under the action of $\xi^\mu$.\\
The Killing vector $\xi^\mu \partial_\mu = \partial_t$ satisfies ($u^\mu = V^{- \frac{1}{2}} \xi^\mu$)
\begin{equation}
u^\mu \partial_\mu = \frac{1}{V} \partial_t\,,\,\,\,u_\mu dx^\mu = - V (dt + Y_i dx^i)
\end{equation}
\begin{equation}
g^{\mu\nu}\partial_\mu \partial_\nu = - \frac{1}{V^2}\, \partial_t^2 + h^{ij}(\partial_i - Y_j dt)(\partial_j - Y_j dt)
\end{equation}
\begin{equation}
(\partial_\mu f^A) dx^\mu = (\partial_i f^A) dx^i\,,\hspace{0.7cm}
H^{AB} = (\partial_i f^A)(\partial_j f^B) h^{ij}
\end{equation}
We add 2 cautionary remarks on rotation:
\begin{itemize}
\item If the Killing vector is rotating, i.e.
\begin{equation}
\omega_{\mu\nu}\,dx^\mu dx^\nu = - V\, D_{[i} Y_{j]} \,dx^i dx^j
\end{equation}
 is non-zero, time independence does not necessarily mean 'stationarity' in the usual sense. Take Minkowski case with $\xi$ the 'helical Killing vector $\xi^\mu \partial_\mu = \partial_t + \omega\, \eta^i \partial_i$, where $\eta$ is a spatial rotation, so the configuration is 'time independent in a rotating frame'.
\item
Secondly, the theory as described breaks down if $f^{- 1}(\mathcal{B})$ is too large in the sense of extending across the 'light cylinder' where $\xi$ becomes null.
\end{itemize}
\section{Relativistic elasticity}
Elasticity is a diffeomorphism invariant theory of configurations on a spacetime $(M,g_{\mu\nu})$, thus defined by an action
\begin{equation}
S [f;g] = \int \rho(f^A, \partial_\mu f^B;g^{\mu\nu})\, \sqrt{- g} \,d^4x
\end{equation}
with the $f^A$'s configurations and with the Lagrangian (called $L = \rho$ for reasons which will become clear shortly) diffeomorphism invariant as described in Sect.2. We next derive an intriguing-looking variational identity which among diffeomorphism invariant theories of maps is specific to elasticity.\\
{\bf{Theorem:}} Let $\delta \rho$ be the first variation of $\rho$ w.r. to $f$, i.e.
\begin{equation}
\delta \rho = \frac{\partial \rho}{\partial f^A}\, \delta f^A + \frac{\partial \rho}{\partial \partial_\mu f^A}\, \partial_\mu \delta f^A\,.
\end{equation}
Then
\begin{equation}\label{remarkable}
\delta \rho - \nabla_\mu(\rho\, \psi^\mu{}_A \delta f^A) = T^\mu{}_\nu \nabla_\mu (\psi^\nu{}_A \delta f^A)\,.
\end{equation}
{\bf{Proof:}} First note that
\begin{equation}\label{First}
\delta \rho = \mathcal{E}_A \delta f^A + \nabla_\mu \left(\frac{\partial \rho}{\partial \partial_\mu f^A}\, \delta f^A\right) = - (\psi^\nu{}_A \nabla_\mu T^\mu{}_\nu)\delta f^A + \nabla_\mu \left(\frac{\partial \rho}{\partial \partial_\mu f^A}\right)\,,
\end{equation}
where (\ref{Bianchi}) was used in the 1st term on the right. Writing
\begin{equation}\label{byparts}
- (\psi^\nu{}_A \nabla_\mu T^\mu{}_\nu)\partial f^A = T^\mu{}_\nu \nabla_\mu (\psi^\nu{}_A \delta f^A) - \nabla_\mu (T^\mu{}_\nu \psi^\nu{}_A \delta f^A)\,,
\end{equation}
using $T^\mu{}_\nu = (\partial_\nu f^A)\frac{\partial \rho}{\partial \partial_\mu f^A} - \rho\, \delta^\mu{}_\nu$ in the 2nd term in (\ref{byparts}) and inserting into (\ref{First}), there is one cancellation, and (\ref{remarkable}) results. This ends the proof.\\
Eq.(\ref{remarkable}) of course entails the result that the EOM are given by $\psi^\mu{}_A \nabla_\nu T^\nu{}_\mu = 0$, which in turn is equivalent to $\nabla_\nu T^\nu{}_\mu = 0$. The total divergence on the l.h. side of (\ref{remarkable}) will play a role when boundary conditions are treated in Sect.9.\\
To further develop the theory as a branch of continuum mechanics, the missing piece of structure is a notion of 'particle number density'. This is done by picking a volume form $\Omega_{ABC}(X)$ on $\mathcal{B}$ and setting
\begin{equation}\label{orient}
(f^\star \Omega)_{\mu\nu\lambda} := J_{\mu\nu\lambda} = (\partial_\mu f^A)(\partial_\nu f^B)(\partial_\lambda f^C) \Omega_{ABC} = n \, \epsilon_{\mu\nu\lambda\sigma}u^\sigma\,,\hspace{0.7cm}n > 0
\end{equation}
Note that $n > 0$ in (\ref{orient}) entails the assumption that $f^A$ preserve orientation. Eq.(\ref{orient}) defines $n$ algebraically in terms of $(f,\partial f;g)$ and is invariant
under diffeomorphisms preserving orientation. Moreover
\begin{equation}
6\, n^2 = H^{AA'}H^{BB'}H^{CC'}\Omega_{ABC\,}\Omega_{A'B'C'}
\end{equation}
and
\begin{equation}
\frac{\partial n}{\partial \partial_\mu f^A} = n \psi^\mu{}_A\,.
\end{equation}
Most importantly, $\nabla_\mu (n u^\mu) = 0$ holds as an identity. To see this recall that the exterior differential $d$ commutes with pull back and note that $(d\Omega)_{ABCD}$ is zero. \\ \\
Taking $\Omega_{ABC} = \varepsilon_{ABC}$, and in a local Lorentzian frame $(t, x^i)$ at $p$ of $M$ where $v^\mu \partial_\mu|_p = (\partial_t + V^i \partial_i)|_p$, there holds
\begin{equation}
n = \mathrm{det}(\partial_i f^A)(1 - V^2)^\frac{1}{2}
\end{equation}
More generally, in the language of Sect.(3.2)
\begin{equation}
n^2 = \mathrm{det}(H^{AB}) = [\mathrm{det}(\partial_k f^B)]^2 \, \mathrm{det}(g^{ij} - \frac{W^i W^j}{N^2})
\end{equation}
The importance of the concept of particle number density will become apparent soon. Furthermore, as opposed to the nonrelativistic theory, particle number carries information about kinetic energy. \\
We now come back to the elastic Lagrangian. From Sect.2 we know that diffeomorphism  invariance implies
\begin{equation}
\frac{\partial \rho}{\partial g^{\mu \nu}} u^\mu = 0
\end{equation}
so $\frac{\partial \rho}{\partial g^{\mu \nu}}$ is 'horizontal'. For example
\begin{equation}
\frac {\partial n}{\partial g^{\mu \nu}} = \frac{n}{2} h_{\mu\nu} \,\,\Longrightarrow\,\,\,2\,\frac {\partial n}{\partial g^{\mu \nu}} - n \, g_{\mu\nu} = n u_\mu u_{\nu}\,.
\end{equation}
In other words, the energy momentum tensor of the Lagrangian $\rho = n$ is 'purely vertical', i.e. $T_{\mu\nu} = n u_\mu u_\nu$ - which is the energy momentum tensor of dust. Thus, if we factorize $\rho$ as
$\rho = n \,\epsilon$, this gives the linear decomposition
\begin{equation}
T_{\mu\nu} = n \epsilon u_\mu u_\nu + 2 n \frac{\partial \epsilon}{\partial g^{\mu\nu}}
\end{equation}
into a vertical 'energy- momentum part' and a horizontal 'stress part'. The latter is of course the relativistic version of (minus) the Cauchy stress tensor.
Furthermore $T^\mu{}_\nu u^\nu = \rho u^\mu$ which justifies the terminology $L = n \epsilon = \rho$, the rest energy density of the elastic material. The quantity $\epsilon$ is the relativistic version of the stored energy function of nonrelativistic elasticity. Note that $T^{\mu\nu}$ can never be zero, provided that $\epsilon > 0$.

Consider the special case where $\rho = n \epsilon (n)$. It turns out that
\begin{equation}\label{fluid}
T^{\mu\nu} = n \epsilon\, u^\mu u^\nu + n^2 \epsilon' h^{\mu\nu}
\end{equation}
i.e. a barotropic fluid with energy density $\rho = n \epsilon$, internal energy $\epsilon$ and pressure $p = n^2 \epsilon'$.
The function $\epsilon(n)$ for fluids and $\epsilon (f^A, H^{BC})$ in general
is determined by the material in question, i.e. plays the role of equation of state - 'stored energy' in classical elasticity. Finding the 'right' EOS for a particular material is a mixture of phenomenology together with ellipticity conditions hopefully guaranteeing good causal properties or existence of equilibria in the time independent case.\\
It might occasionally be useful to change the Lagrangian by changing the volume form $\Omega_{ABC}$ for fixed stored energy $\epsilon$. Here is how the EOM changes under $\overline{L} = \bar{\Omega}(f) L$:
\begin{equation}
\overline{\mathcal{E}}_A = \overline{\Omega}\,\mathcal{E}_A - \left[(\partial_\nu f^B) \frac{\partial \rho}{\partial_\nu f^A} - \rho \,\delta_A{}^B\right]\partial_B \overline{\Omega}
\end{equation}
\vspace{0.2cm}
The quantity $M^{\mu\nu}_{AB} = \frac{\partial^2 \rho}{\partial (\partial_\mu f^A)\partial (\partial_\nu f^B)}$, similarly to $T_{\mu\nu}$, will have a vertical part, determining the leading-order $\partial_t^2$ term in the EOM plus a horizontal part containing the spatial derivatives.
In more detail (setting $\tau_{AB} = 2\, \frac{\partial \epsilon}{\partial H^{AB}}$)
\begin{equation}
T^{\mu\nu} = n \epsilon u^\mu u^\nu + n \, \tau_{AB}(\partial_\sigma f^A)(\partial_\lambda f^B)g^{\mu\sigma}g^{\nu\lambda}\,,
\end{equation}
then $M^{\mu\nu}_{AB} = M^{\nu\mu}_{BA}$ is given by
\begin{equation}
M^{\mu\nu}_{AB} = - n (\epsilon\, H_{AB} + \tau_{AB}) u^\mu u^\nu + U_{ACBD}(\partial_\sigma f^C)(\partial_\lambda  f^D)g^{\mu\sigma}g^{\nu\lambda}
\end{equation}
\\
\vspace{0.3cm}
where $U_{ACBD} = U_{BDAC}$ satifies
\begin{equation}
U_{ACBD} = n\left(\tau_{AB}H_{CD} + \tau_{AC}H_{BD} + \tau_{BD}H_{AC} + 2 \, \frac{\partial \tau_{AB}}{\partial H^{CD}} + 2 \,\epsilon H_{A[C}H_{D]B}\right)
\end{equation}
Note $M^{\mu\nu}_{AB}$ depends on $(f, \partial f)$, so
\begin{equation}
- \mathcal{E}_A = 0 = M^{\mu\nu}_{AB}\partial_\mu \partial_\nu f^B + \mathrm{l.o.}
\end{equation}
is a 2nd order quasilinear system of PDE's for $f^A$.
\\

The local initial value problem in the analytic case with initial surface $\Psi (x) = 0$ is solvable iff $M^{\mu\nu}_{AB} (\partial_\mu \Psi)(\partial_\nu \Psi)$ is invertible.
This can certainly be achieved when $|\tau_{AB}|$ and $|\frac{\partial \tau_{AB}}{\partial H^{CD}}|$ are both $\ll \epsilon$ and $|H^{AB} - \delta^{AB}|$ is small. Note in particular that conditions on $T_{\mu\nu}$ like the dominant energy condition do not imply wellposedness - which is of course known already for perfect fluids, since sound speed needs 2nd derivatives of $\epsilon$.\\
Finally in this section we turn to the issue of simple solutions, which do not require solving any partial differential equations. In some field theories such are easy to find, typically 'vacuum' solutions e.g. spacetimes with constant curvature as solutions for the Einstein vacuum equations with cosmological constant. In the case of elasticity, even in the background of Minkowski spacetimes, this requires assumptions on the stored energy function. \\
As an example consider $(M,g)$ an ultrastatic spacetime, i.e. with a timelike Killing vector which is hypersurface orthogonal and geodesic, so that in suitable coordinates
\begin{equation}
g = - dt^2 + h_{ij}(x) dx^i dx^j
\end{equation}
For the body we take $(\mathcal{B}, G_{AB})$, where $G_{AB}(X) dX^A dX^B$  is isometric to $h_{ij}$, that is there exists a diffeomorphism $\mathring{f}: x^i \mapsto X^A = \mathring{f}^A(x)$, s.th.
\begin{equation}
G^{AB}(\mathring{f}(x)) = (\partial_i \mathring{f}^A)(x)(\partial_j \mathring{f}^A)(x)h^{ij}(x)\hspace{0.4cm}\mathrm{or}\hspace{0.4cm}G_{AB} = (\mathring{\Phi}^\star h)_{AB}\,,
\end{equation}
where $\mathring{\Phi}^\star$ is pull-back under $\mathring{\Phi} = (\mathring{f})^{- 1}$, and for the volume element on $\mathcal{B}$ we take $\sqrt{\mathrm{det}G} \,\varepsilon_{ABC}$.\\
Clearly $f^A(t,x) = \mathring{f}^A(x)$ is an allowable, in fact: non-contorted, configuration with
\begin{equation}
\mathring{H}^{AB}(t,x) = G^{AB}(\mathring{f}(x))
\end{equation}
Next require that the function $\epsilon (f^A, H^{BC})$ satisfies
\begin{equation}
\frac{\partial \epsilon(X, H^{CD})}{\partial H^{AB}}|_{H^{CD} = G^{CD}(X)} = 0
\end{equation}
Then $\mathring{f}$ solves the field equations. To prove this assertion, first write
\begin{equation}
\mathring{T}^{\mu\nu} = \mathring{n} \mathring{\epsilon} u^\mu u^\nu + \mathring{t}^{\mu\nu}\,,
\end{equation}
where $u^\mu \partial_\mu = \partial_t$ and $\mathring{t}^{\mu\nu} \partial_\mu \partial_\nu = 2 \mathring{n} \frac{\partial \epsilon (\mathring{f},H^{CD})}{\partial H^{AB}}|_{H^{CD} = \mathring{H}^{CD}} (\partial_i \mathring{f}^A) (\partial_j \mathring{f}^B) h^{ik} h^{jl}\,\partial_k \partial_l$ is clearly zero. Then note that $u^\mu \partial_\mu \mathring{H}^{AB} = 0$, whence $\mathring{n}$ and $\mathring{\epsilon}$ are also invariant under $u^\mu$. Furthermore $u^\mu$ is geodesic. Thus the first term in $\mathring{T}^{\mu\nu}$ has zero divergence, which ends the proof.

\section{Elasticity \`a la Tahvildar-Zadeh}

As we have seen in the last section, perfect fluids are a special case of our equations. However those are usually, and preferably, described in terms of the quantities 4-velocity and  energy density as function of pressure, all viewed as fields on $M$, which obey a 1st order system of PDE's. We repeat that in our present framework the 4-velocity is a derived quantity, namely an algebraic function of the partial derivatives of the configuration and of the spacetime metric. Still one may ask if there is a formulation of elasticity in the spirit of hydrodynamics, from which configurations whence the material manifold have disappeared. In this section we describe such a formulation, due to Tahvildar-Zadeh \cite{Tahvildar1998relativistic}, corresponding to elasticity in the so-called isotropic case. Here the $3 \times 4 = 12$ partial derivatives of configurations are replaced by the 3+6=9 independent quantities given by 4-velocity $u^\mu$ and a symmetric tensor $\gamma_{\mu\nu}$ describing 'strain', constrained by the conditions $u_\mu u^\mu = - 1$ and $\gamma_{\mu\nu} u^\nu = 0$, all subject to a 1st order system of PDE's. Missing from this formulation are the 3 degrees of freedom corresponding to rotations.\\
Suppose $\rho$ depends only on eigenvalues of $H^{AB}$ w.r. to some a priori given metric $G_{AB}(X)$ on $\mathcal{B}$ (isotropic case). The pull-back of $G$ under $f$,
\begin{equation}
\gamma_{\mu\nu}(x) = (\partial_\mu f^A)(x)(\partial_\nu f^B)(x)\, G_{AB}(f(x))
\end{equation}
by fiat satisfies
\begin{equation}\label{fiat}
\mathcal{L}_u \gamma_{\mu\nu} = 0\,,\,\hspace{1.3cm}\gamma_{\mu\nu} u^\mu = 0
\end{equation}
and the non-zero eigenvalues of $\gamma_\mu{}^\nu$ are identical with the eigenvalues of $H_A{}^B = G_{AC} H^{BC}$. By the Cayley-Hamilton theorem $T_\mu{}^\nu$ is a linear combination of $(1,1)$-tensors $u_\mu u^\nu, h_\mu{}^\nu, \gamma_\mu{}^\nu, \gamma^2_\mu{}^{\,\nu}$ with coefficients depending only on the eigenvalues of $\gamma_\mu{}^\nu$. Field equations are $\nabla_\nu T_\mu{}^\nu = 0$ together with (\ref{fiat}) and the condition $u^\mu u_\mu = - 1$. The identity
$u_\mu \nabla_\nu T_\mu{}^\nu = 0$ will of course still hold, but particle number conservation needs checking. Namely there holds
\begin{equation}
6 \,\gamma_{\mu[\mu'}\gamma_{|\nu|\nu'}\gamma_{|\rho|\rho']} = J_{\mu\nu\rho}\,J_{\mu'\nu'\rho'}
\end{equation}
Thus
\begin{equation}
(\mathcal{L}_u J_{\mu\nu\rho})\,J_{\mu'\nu'\rho'} + J_{\mu\nu\rho}\,(\mathcal{L}_u J_{\mu'\nu'\rho'}) = 0
\end{equation}
with $J_{\mu\nu\rho}$ nowhere zero. This further implies $\mathcal{L}_u J_{\mu\nu\rho} = 0$. But
\begin{equation}
\mathcal{L}_u J = u \lrcorner \,d J + d (u \lrcorner \,J) =  u \lrcorner \,d J
\end{equation}
Thus, since $d J$ is a 4-form, $d J = 0$, which we set out to prove. Thus we have a total of 6 independent 1st order equations for the 6 quantities $(u^\mu, \gamma_{\nu\rho})$, that is a closed system from which the material manifold has disappeared. In \cite{Tahvildar1998relativistic} it is shown that, under reasonable conditions on the stored energy, the system is weakly hyperbolic in the sense that the characteristic speeds are real, i.e. the characteristic polynomial is hyperbolic in the sense of Sect.8. Unfortunately this does not always imply wellposedness. On the other hand, given a solution with data on some spacelike hypersurface $\Sigma$
so that the pull back to $\Sigma$ of $\gamma_{\mu\nu}$ is the pull back of $G_{AB}$ under some map $\bar{f}: \Sigma \rightarrow \mathcal{B}$, we can reconstruct the configuration $f^A$ by solving $u^\mu \partial_\mu f^A = 0$ with initial data $f^A|_\Sigma = \bar{f}^A$.
\section{Nonrelativistic limit}
Take $(M,g)$ to be Minkowski spacetime $(\mathbb{R}^4,\eta_{\mu\nu})$ with $\eta^{\mu\nu}\partial_\mu \partial_\nu = - \frac{1}{c^2}(\partial_t)^2 + \delta^{ij}\partial_i \partial_j$ and take
\begin{equation}
\epsilon = m_0 c^2 + W (f,H^{AB}) = m_0 c^2 + W (f, K^{AB}) + O(\frac{1}{c^2})\,,
\end{equation}
where $m_0$ is (rest-mass)/(particle)(volume). We set $m_0 = 1$ from now on. Furthermore
\begin{equation}
H^{AB} = (\partial_i f^A)(\partial_j f^B)\left(\delta^{ij} - \frac{V^i V^j}{c^2}\right)
\end{equation}
 with $K^{AB} = (\partial_i f^A)(\partial_j f^B) \delta^{ij}$ the so-called inverse Cauchy-Green strain tensor. $W$ is called stored energy \footnote{This $W$ should not be confused with $W$ appearing in $W^2 = W^i W_i$ in Sect.(3.1)}and theories of this type are called 'hyperelastic, frame-indifferent'. Hyperelasticity means that the theory is derivable from a Lagrangian, frame indifference corresponds in our setting to diffeomorphism invariance.\\
Furthermore (recall from (\ref{recall}) that $V^i = - \phi^i{}_A \,\partial_t f^A$)
\begin{equation}
n = (\mathrm{det} H^{AB})^\frac{1}{2} = \mathrm{det}(\partial_i f^A)\left(1 -  \frac{V^2}{c^2}\right)^\frac{1}{2}
\end{equation}
\begin{equation}
S[f] = \int \mathrm{det}(\partial_i f^A)(c^2 - \frac{V^2}{2} + W)\, dt\, d^3x + O(\frac{1}{c^2})
\end{equation}
Note: the divergent term is a Null Lagrangian, i.e. the EL-equations for that term are identically satisfied, i.e.
\begin{equation}
\partial_i [\mathrm{det}(\partial_j f^B) \phi^i{}_A] = 0\,,
\end{equation}
which is the famous Piola identity.

The 'renormalized' Lagrangian, i.e. with the divergent term omitted, breaks Galilean invariance, but not of course the EOM.
In fact the Galilean equations have a spacetime form. The definition of the particle number density, now called $k$, is as before is
\begin{equation}
(\partial_\mu f^A)(\partial_\nu f^B)(\partial_\lambda f^C) \Omega_{ABC} = k\, \epsilon_{\mu\nu\lambda\sigma}u^\sigma\,,\hspace{0.7cm}n > 0
\end{equation}
or $k = \mathrm{det}(\partial_i f^A)$, when $\Omega_{ABC} = \varepsilon_{ABC}$. Note $k$ contains no time derivatives.
The stress-energy tensor $T^{\mu\nu}$ is then given by
\begin{equation}
T^{\mu\nu} =  k v^\mu v^\nu + 2\, k \,\frac{\partial W}{\partial H^{AB}}(\partial_\rho f^A) (\partial_\sigma f^B)e^{\mu\rho} e^{\nu\sigma}
\end{equation}
Then $\partial_\mu (k v^\mu) = 0$ and $\tau_\mu \partial_\nu T^{\mu\nu} = 0$ hold as identities, and the 3
EOM are $\partial_\nu T^{\mu\nu} = 0$.\\
In the presence of a gravitational potential $U$ we define the 'gravitational stress' by
\begin{equation}
\Theta^{\mu\nu} = \frac{1}{4 \pi G}(e^{\mu\rho}e^{\nu\sigma}- \frac{1}{2}e^{\mu\nu}e^{\rho\sigma})(\partial_\rho U)(\partial_\sigma U)
\end{equation}
Then
\begin{equation}
\partial_\nu \Theta^{\mu\nu} = \frac{1}{4 \pi G} (e^{\mu\nu} \partial_\nu U) \,\Delta U\,\,\mathrm{where}\,\,\Delta = h^{\mu\nu} \partial_\mu \partial_\nu\,,
\end{equation}
and the coupled system of equations can concisely be written as
\begin{equation}
\partial_\nu (T^{\mu\nu} + \Theta^{\mu\nu}) = 0\,,\hspace{0.5cm}\Delta U = 4 \pi G \,n
\end{equation}
\section{Linearization at a stressfree state}
We assume for simplicity that the material is homogenous and isotropic, so the equations of motion are
\begin{equation}
- \mathcal{E}_A = M^{\mu\nu}_{AB}\nabla_\mu \partial_\nu f^B = 0
\end{equation}
Here: we linearize at a stressfree configuration in Minkowski, i.e. $\eta^{\mu\nu}\partial_\mu \partial_\nu = - \frac{1}{c^2}\partial_t^2 + \delta^{ij}\partial_i \partial_j$ and take
\begin{equation}
\mathring{f}^A = \delta^A{}_i x^i\,\,\Rightarrow \partial_\mu \partial_\nu \mathring{f}^A = 0\,\,\Rightarrow\,\,\mathring{\mathcal{E}}_A = 0
\end{equation}
and the linearized equations are of the form
\begin{equation}
\mathring{M}^{\mu\nu}_{AB}\,\partial_\mu \partial_\nu \,\delta f^B = 0
\end{equation}
Further constitutive assumptions are:
\begin{itemize}
\item $\epsilon|_{H = \delta} = c^2$
\item $\frac{\partial \epsilon}{\partial H^{AB}}|_{H = \delta} = 0$
\item $\frac{\partial^2 \epsilon}{\partial H^{AB}\partial H^{CD}}|_{H = \delta} = \frac{1}{4} (q \,\delta_{AB}\delta_{CD} + 2 r \, \delta_{C(A}\delta_{B)D})$ \hspace{0.4cm}$r, q$\,\,\,constants
\end{itemize}
So (note $\mathring{n} = 1,\,\mathring{H}^{AB} = \delta^{AB}\,,\,\,\mathring{u}^\mu = \frac{1}{c}\, \delta^\mu{}_0\,,\,\,\,\mathring{\tau}^{AB} = 0$)
\begin{equation}
\mathring{M}^{\mu\nu}_{AB} = [- \delta_{AB}\, \delta^\mu{}_0 \delta^\nu{}_0 + (q + r)\, \delta^\mu{}_A \delta^\nu{}_B + r \,\delta^\nu{}_A \delta^\mu{}_B]\,,
\end{equation}

so that
\begin{equation}
\delta \mathcal{E}_i = - \partial_t^2 \,\delta\! f_i + \partial_j \,(L_{ijkl}\, \partial_{(k}\delta f_{l)})\,,
\end{equation}
where $L_{ijkl} = L_{(ij)(kl)} = L_{klij}$ is given by
\begin{equation}
L_{ijkl} = q \,\delta_{ij} \delta_{kl} + 2 r \,\delta_{i(k}\delta_{l)j}
\end{equation}

The following ellipticity conditions for the spatial operator play a role:\\
(i) $L_{ijkl}\lambda_i \mu_j \lambda_k \mu_l > 0$: this is called rank-one positivity or the Legendre-Hadamard condition. In the present case this means $r = c_2^2 > 0,\,2 r + q = c_1^2 > 0$. It then follows that the characteristic polynomial $P(k)$ is given by (setting $c = 1$)
\begin{equation}
\mathrm{det} (\mathring{M}^{\mu\nu}_{AB}k_\mu k_\nu) = (g_1^{\mu\nu}k_\mu k_\nu)(g_2^{\mu\nu}k_\mu k_\nu)^2\,,
\end{equation}
\begin{equation}
g_1^{\mu\nu} = \eta^{\mu\nu} + \mathring{u}^\mu \mathring{u}^\nu \left(1 - \frac{1}{c_1^2}\right)\,,\hspace{0.6cm}g_2^{\mu\nu} = \eta^{\mu\nu} + \mathring{u}^\mu \mathring{u}^\nu \left(1 - \frac{1}{c_2^2}\right)
\end{equation}
The $g_1$ - cone determines the phase speed of longitudinal sound waves, the $g_2$ one the tranversal ('shear') waves.\\ \\
(ii) $L_{ijkl}m_{ij} m_{kl} > 0$: 'pointwise stability'.  This requires $r > 0,\,\frac{2 r}{3} + q > 0$. \\ \\
(ii) plays the following role. Consider the Neumann-type boundary value problem on $\mathbb{R}^3$ given by
\begin{equation}
\partial_j (L_{ijkl}\,\partial_{(k}\,\delta f_{l)}) = b_i\,\,\,\,\mathrm{in}\,\,\mathring{f}^{- 1}(\mathcal{B}) \in \mathbb{R}^3,\hspace{0.8cm}
L_{ijkl}\,\partial_{(k}\,\delta f_{l)}\,n_j|_{\mathring{f}^{- 1}(\partial \mathcal{B})} = \tau_i
\end{equation}
Then, in appropriate function spaces, this has the 'right' kernel and cokernel, namely kernel spanned by Euclidean Killing vectors $\delta f_i = \xi_i$, and the range spanned by
\begin{equation}\label{corange}
\int_{\mathring{f}^{- 1}(\mathcal{B})} \xi^i b_i \,d^3\!x + \int_{\mathring{f}^{- 1}(\partial \mathcal{B})} \xi^i \tau_i \, dS(x) = 0
\end{equation}
In the language of nonrelativistic elasticity: 'load $b_i$ and surface traction $\tau_i$ have to be equilibrated', see Sect.11.
\section{Elastic causality, wellposedness}
There is, for equations of the form
\begin{equation}
M^{\mu\nu}_{AB}[f,\partial f;x]\,\partial_\mu \partial_\nu f^B + \,\mathrm{l.o.} = 0
\end{equation}
with $M^{\mu\nu}_{AB} = M^{\nu\mu}_{BA}$ a theory of local wellposedness by Hughes, Kato, Marsden \cite{hughes1977well} and in geometrical form by Christodoulou in \cite{christodoulou2000action}:\\
{\bf{Definition:}} The partial differential operator $M^{\mu\nu}_{AB} \,\partial_\mu\partial_\nu f^B + \mathrm{l.o.}$ is called regularly hyperbolic w.r. to $(\xi_\mu, X^\nu)$, if there exists a 'subcharacteristic' covector $\xi_\mu$ and a 'timelike' vector $X^\mu$ with $X^\mu \xi_\mu > 0$, s.th. \\
(i)$\,\,\,M^{\mu\nu}_{AB} \,\xi_\mu \xi_\mu$ is negative definite.\\
(ii) $M^{\mu\nu}_{AB} \lambda^A \lambda^B \eta_\mu \eta_\nu > 0$ for all $\eta_\mu$ with $X^\mu \eta_\mu = 0$.\\ \\
Our case as above is regular hyperbolic w.r. to $(\xi_\mu = - u_\mu, X^\mu = u^\mu)$ provided  $\epsilon\, H_{AB} + \tau_{AB}$ is positive definite and $U_{ACBD}$ is rank-1 positive.
The set of pairs $(\xi_\mu, X^\nu)$ consists of 2 connected components of opposite pairs $(\xi^+, X^+) \cup (\xi^{-}, X^{-})$ all of which are convex.\\
The connection with the characteristic polynomial $P(k) = \mathrm{det} (M^{\mu\nu}_{AB} \,k_\mu k_\nu)$ is as follows. A homogenous polynomial
in $k_\mu$  is called a hyperbolic polynomial (Garding \cite{gaarding1959inequality}) w.r. to a covector $\xi_\mu$
 if the equation
 \begin{equation}
P(\eta + \lambda \xi) = \mathrm{det}[M^{\mu\nu}_{AB}(\eta_\mu + \lambda \xi_\mu)(\eta_\nu + \lambda \xi_\nu)] = 0\hspace{1cm}
\end{equation}
has only real (in our case $2 \times 3 = 6$) solutions $\lambda$ corresponding to the 6 sheets of the characteristic set $\mathfrak{C}$. That can in general be  complicated, have singularities and is perhaps ultimately a matter for algebraic geometry.\\
For a regularly hyperbolic operator the set of $\xi$'s turns out to be equal to the union $\{\xi^+\}\cup\{\xi^-\}$.
For our previous example 2 pairs of these solutions cones coincide.\\
 Note: for the $\xi^+$'s the faster ($c_1 > c_2$) cone lies inside the slower one. This is because the covectors $\xi$ describe 'acoustically spacelike' hypersurfaces, i.e. hypersurfaces foliating a domain of dependence, which is the smaller the faster the phase velocity. The $X$'s are dual objects describing 'rays', i.e. domains of influence. The faster the phase speed, the larger the domain of influence. Formally
 \begin{equation}
 g_{1,2}^{\mu\nu} = \eta^{\mu\nu} + u^\mu u^\nu(1 - \frac{1}{c_{1,2}^2})\,,\hspace{0.4cm} g_{\mu\nu}^{1,2} = \eta_{\mu\nu} + u_\mu u_\nu (1 - c_{1,2}^2)
 \end{equation}
 In nonrelativistic elasticity the same relations hold after replacing $\eta^{\mu\nu} + u^\mu u^\nu$ with the Galilean $h^{\mu\nu}$ and $u^\mu$ by $v^\mu$ with $v^\mu \tau_\mu = 1$ and
 \begin{equation}
 g_{\mu\nu}^{1,2} = h_{\mu\nu} - \tau_\mu \tau_\nu\, c_{1,2}^2\,\,\,\,\mathrm{with}\,\,\,\,h^{\mu\nu} h_{\nu\rho} = \delta^\mu{}_\rho - v^\mu \tau_\rho
 \end{equation}
Christodoulou in \cite{christodoulou2000action} proved a uniqueness theorem for regularly hyperbolic equations in domains of dependence foliated by spacelike surfaces. For existence one should be able to  use the results of \cite{hughes1977well}.\\
As for the  connection with symmetric hyperbolicity (Beig \cite{beig2006concepts}): replace
\begin{equation}
M^{\mu\nu}_{AB} \,\partial_\mu\partial_\nu f^B = \mathrm{l.o.}
\end{equation}
with
\begin{equation}
W^{\mu\nu}{}_{AB}{}^{(\sigma)}\partial_\sigma F^B{}_\nu = \mathrm{l.o.}\hspace{1cm}w^\mu \partial_\mu f^A = F^A{}_\mu w^\mu
\end{equation}
\begin{equation}
W^{\mu\nu}{}_{AB}{}^{(\sigma)} = w^\mu M^{\sigma\nu}_{AB} + 2 w^{[\nu} M^{\sigma]\mu}_{BA}
\end{equation}
for a suitable vector $w^\mu$.
Clearly $W^{\mu\nu}{}_{AB}{}^{(\sigma)} = W^{\nu\mu}{}_{BA}{}^{(\sigma)}$, so the new system is symmetric. Sometimes, when $w^\mu = X^\mu$, $W^{\mu\nu}{}_{AB}{}^{(\rho)} \xi_\rho$ is positive definite, but that is not guaranteed, e.g. for elasticity this requires pointwise stability.
We remark that none of the above works for fluids since the 'spacelike' term is degenerate there.\\
A local wellposedness result for elasticity coupled to gravity should follow the standard path (see e.g. Ringstr\"om \cite{ringstrom2009cauchy}), but a detailed presentation seems to be lacking. Whether the gravitational cone is the fastest one or not should play no role, when only local wellposedness is concerned.\\
Finally let us mention that for nonrelativistic elasticity there are even global results on $\{t\} \times \mathbb{R}^3$ for certain equations of state which lead to a null condition (see Sideris \cite{sideris1996null, sideris2000nonresonance}).
\section{Bodies surrounded by vacuum}
When $\mathcal{B}$ is compact, the Lagrangian density can only be integrated over $f^{- 1}(\mathcal{B})$, which is the infinite region of spacetime 'inside' the timelike hypersurface $f^{- 1}(\partial \mathcal{B})$.
We write
\begin{equation}\label{basic1}
S [f] = \int_{f^{- 1}(\mathcal{B})}\!\!\!\! \rho \,\sqrt{- g} \,d^4x = \int_{f^{- 1}(\mathcal{B})}\!\!\!\! \rho  \,dV_g\,,
\end{equation}
where we assume that $f^{- 1}(\mathcal{B})$ is cut off in the future and past by two non-intersecting spacelike hypersurfaces $\Sigma_1$ and $\Sigma_2$.  Suppose we vary (\ref{basic1}) w.r. to $f$ with the understanding that the $f^A$'s are all equal to a given map outside a compact proper sub-region which however may include $f^{- 1}(\partial \mathcal{B})$. Then we have the\\
{\bf{Theorem:}} There holds the identity
\begin{equation}\label{basic2}
\delta S = \int_{f^{- 1}(\mathcal{B})}\!\!\! T^\mu{}_\nu \nabla_\mu (\psi^\nu{}_A \delta f^A)\,dV_g\,,
\end{equation}
where $\delta f^A$ has 'compact support in time', i.e. vanish on $\Sigma_1$ and $\Sigma_2$, but need not satisfy any further condition on $f^{- 1}(\partial \mathcal{B})$. \\
{\bf{Proof:}} Using the identity (\ref{remarkable}) together with the formula
\begin{equation}
\delta \int_{f^{- 1}(\mathcal{B})}\!\!\! dV_g = - \int_{f^{- 1}(\partial \mathcal{B})}\!\!\psi^\mu{}_A \delta f^A d\Sigma_\mu\,,
\end{equation}
the contribution from the total divergence in (\ref{remarkable}) drops out, and we get (\ref{basic2}). Alternatively one can prove (\ref{basic2}) directly from diffeomorphism invariance of (\ref{basic1}).\\  \\
Suppose now we ask for an extremum of the action $S[f]$ subject only to the condition that variations have compact support in time.
This is similar, in the calculus of variations (see e.g. \cite{giaquinta1995calculus}), to asking for 'natural boundary conditions'. Then, using the identity (\ref{basic1}), it is easy to see that this implies both the Euler-Lagrange equation
\begin{equation}\label{bulk}
\psi^\mu{}_A \nabla_\nu T_\mu{}^\nu = 0\hspace{0.4cm}\mathrm{in}\hspace{0.1cm}f^{- 1}(\mathcal{B})
\end{equation}
and the boundary condition
\begin{equation}\label{boundary}
\psi^\mu{}_A T_\mu{}^\nu m_\nu|_{f^{- 1}(\partial\mathcal{B})} = 0\hspace{0.2cm}\Leftrightarrow \hspace{0.2cm}\frac{\partial \epsilon}{\partial \partial_\mu f^A}\,
m_\mu|_{f^{- 1}(\partial\mathcal{B})} = 0
\end{equation}
to be satisfied, where $m_\mu$ is the (spacelike) conormal of $f^{- 1}(\partial\mathcal{B})$. Recall that the vertical component of $\nabla_\nu T_\mu{}^\nu$ namely $u^\mu \nabla_\nu T_\mu{}^\nu$ is identically zero. Similarly the vertical component of $T_\mu{}^\nu m_\nu|_{f^{- 1}(\partial\mathcal{B})}$, namely  $u^\mu T_\mu{}^\nu m_\nu|_{f^{- 1}(\partial\mathcal{B})}$ also vanishes due to (\ref{mixed}) since $u^\mu m_\mu = 0$. \\
Summing up, we take as requirement on the matter-vacuum interface the condition that $T_\mu{}^\nu m_\nu|_{f^{- 1}(\partial\mathcal{B})}$ be zero with the idea that this corresponds to bodies 'floating freely in spacetime'. This is supported by the following fact: let the spacetime $(M,g_{\mu\nu})$ have a Killing vector $\xi^\mu$. Then the integral
\begin{equation}
E_\xi = \int_{\Sigma \cap f^{- 1}(\mathcal{B})}\xi^\mu T_\mu{}^\nu d\Sigma_\nu
\end{equation}
does not depend on the choice of spacelike slice $\Sigma$.\\
Let us point out that $T_\mu{}^\nu m_\nu|_{f^{- 1}(\partial\mathcal{B})} = 0$ is also natural when elasticity is coupled to gravity, i.e.
\begin{equation}
G_{\mu\nu} = \kappa \,T_{\mu\nu}\,\chi_{f^{- 1} (\mathcal{B})}\,,
\end{equation}
where $\chi_{f^{- 1} (\mathcal{B})}$ is the characteristic function of $f^{- 1}(\mathcal{B})$. Now for elastic solids $T_{\mu\nu}$ necessarily drops to zero sharply at the matter-vacuum interface. For regularity it is sensible to require that the first and second fundamental for the metric change continuously across $f^{- 1}(\partial \mathcal{B})$. But since $G_\mu{}^\nu m_\nu|_{f^{- 1}(\partial \mathcal{B})}$, as is well known, contains only up to 1st normal derivatives of the metric and $G_{\mu\nu}$ is zero in the vacuum region, it follows that $G_\mu{}^\nu m_\nu|_{f^{- 1}(\partial \mathcal{B})}$ has to vanish, in accordance with (\ref{boundary}).\\

\section{Material picture}
Trying to solve the elastic equations in the presence of the 'free' boundary   $f^{- 1}(\partial\mathcal{B})$ makes it tempting (almost obligatory?) to rewrite these equations in terms of the inverse of
$f^A$ relative to some slicing of $M$ by spacelike hypersurfaces. We first write the action in terms of the ADM variables introduced previously.
\begin{equation}
\sqrt{- g}\, d^4 x = N [\mathrm{det}(g^{ij} - \frac{Y^i Y^j}{N^2})]^{- \frac{1}{2}} dt d^3x
\end{equation}
We had
\begin{equation}
H^{AB} = (\partial_i f^A)(\partial_j f^B)\, (g^{ij} - \frac{W^iW^j}{N^2})
\end{equation}
so that
\begin{equation}
n^2 = \mathrm{det}(H^{AB}) \Omega^2(f) = [\mathrm{det}(\partial_k f^A)]^2 \, \mathrm{det}(g^{ij} - \frac{W^i W^j}{N^2}) \Omega^2(f)
\end{equation}
Thus
\begin{multline}
S = \int_{f^{- 1}(\mathcal{B})}\!\!\!\! \epsilon(f^A, H^{BC})\, \sqrt{\frac{N^2 - W^2}{N^2 - Y^2}} \,N \,\Omega(f)\,\mathrm{det}(\partial_i f^A)\,dt d^3x =\\
\!\!\!\!\!\!\!\!\!\!\!\!\!\!\!\!\!\!\!\!\!\!\!\!\!\!\!\!\!\!\!\!= \int_{\{T\}\times \mathcal{B}}\!\!\!\epsilon(X, \overline{H}^{AB})\,\sqrt{\frac{\overline{N}^2 - \overline{W}^2}{\overline{N}^2 - \overline{Y}^2}}\,\,\overline{N}\,\Omega(X)\, dT d^3X\,,
\end{multline}
where an overbar denotes composition with $\Phi^i$ defined by
\begin{equation}
f^A(T, \Phi(T,X)) = X^A
\end{equation}

\begin{equation}
\overline{H}^{AB} = \rho^A{}_i \rho^B{}_j \left(\overline{g}^{ij} - \frac{\overline{W}^i\overline{W}^j}{\overline{N}^2}\right)\,,\hspace{0.5cm}
\rho^A{}_j \,\partial_A \Phi^i = \delta^i{}_j
\end{equation}
Note that $\overline{V}^i = \partial_T \Phi^i$. Thus the 'prefactor' in the material action has no space derivatives. Observe metric components are pulled back under $\Phi$ as scalars.
So
\begin{equation}
L_{mat} = \epsilon(X, \overline{H}^{AB})\,\sqrt{\frac{\overline{N}^2 - \overline{W}^2}{\overline{N}^2 - \overline{Y}^2}}\,\,\overline{N}\,\Omega(X)
\end{equation}
is the material Lagrangian and the material EOM are
\begin{equation}\label{bulk1}
- \partial_A\, \left(\frac{\partial L_{mat}}{\partial (\partial_A \Phi^i)}\right) - \partial_T \, \left(\frac{\partial L_{mat}}{\partial (\partial_T \Phi^i)}\right) + \frac{\partial L_{mat}}{\partial \Phi^i} = 0
\end{equation}
with boundary conditions
\begin{equation}\label{boundary1}
\left(\frac{\partial \epsilon}{\partial (\partial_A \Phi^i)}\right)n_A \big{\vert}_{\partial \mathcal{B}} = 0\,,
\end{equation}
which are of course the equivalent of (\ref{bulk},\ref{boundary}) in the material picture.\\
Local wellposedness follows from a theorem by Koch \cite{koch1993mixed}, provided the initial configuration satifies the material version of the previous hyperbolicity condition, involving e.g. $\frac{\partial^2 \epsilon}{\partial (\partial_A \Phi^i)\partial (\partial_B \Phi^j)}$ (see also \cite{wernig2006relativistic} and \cite{beig2007motion}).
\section{Time independent problems}
Suppose $(M;g)$ has a timelike Killing vector $\xi^\mu$. Recall the quotient space $Q$ is endowed with the metric $h_{ij}$. From (\ref{geroch}) it follows that
\begin{equation}
\sqrt{- g} \,d^4x = V dt \,\sqrt{h} \,d^3x\,,
\end{equation}
where $- V^2 = g_{\mu\nu} \xi^\mu \xi^\nu$. So the reduced action for time independent configurations $f^A (t,x^i) = f^A(x^i)$ is
\begin{equation}
S[f;h^{ij},V] = \int_{f^{- 1}(\mathcal{B})} \rho\, V\, \sqrt{h} \,d^3x\,,
\end{equation}
where $\rho$ is function of $(f^A, H^{BC} = (\partial_i f^B) (\partial_j f^C) h^{ij})$.
Note that the presence of the potential $V$ means that the diffeomorphism invariance of the action is now broken. In fact, let $\Phi$ be a diffeomorphism from $Q$ into itself. Then
\begin{equation}
S[f \circ \Phi;\Phi^\star h, V] = S[f;h,V \circ \Phi^{- 1}]
\end{equation}
So
\begin{equation}
- \int_{f^{- 1}(\mathcal{B})}\!\!\! \rho\,\mathcal{L}_\eta V\,\sqrt{h} \,d^3x = \frac{1}{2}\int_{f^{- 1}(\mathcal{B})}\!\!\! t_{ij} \mathcal{L}_\eta h^{ij}\, \sqrt{h} \,d^3x
 + \frac{d}{d \epsilon}\vert_{\epsilon = 0}\, S[f + \epsilon \mathcal{L}_\xi f;g]\,,
\end{equation}
where $t_{ij} = 2 \frac{\partial \rho}{\partial h^{ij}} - \rho h_{ij}$.
Setting again $\eta^i = \psi^i{}_A \delta f^A$, where $\psi^i{}_A \,\partial_i f^B = \delta^B{}_A$,  we find that
\begin{equation}
\delta S = \int_{f^{- 1}(\mathcal{B})} V\,t_{ij} D^i (\psi^j{}_A \delta f^A)\sqrt{h} \,d^3x - \int_{f^{- 1}(\mathcal{B})} \!\!\rho \,D_i V \psi^i{}_A \delta f^A\,\sqrt{h} \,d^3x\,,
\end{equation}
where $D_i$ is the covariant derivative associated with $h_{ij}$.
So the field equations plus boundary conditions are
\begin{equation}\label{reduced}
D_j (V t_i{}^j) + \rho\, D_i V = 0\hspace{0.3cm}\mathrm{in}\hspace{0.1cm}f^{- 1}(\mathcal{B})\,,\hspace{0.6cm}t_i{}^j n_j|_{f^{- 1}(\mathcal{\partial B})} = 0\,.
\end{equation}
Note the 'force term' $\rho D_i V$ in (\ref{reduced}) is zero if and only if the Killing vector $\xi^\mu$ is geodesic.\\
As an example consider Minkowski spacetime $(M, \eta_{\mu\nu})$ and $\xi^\mu\partial_\mu = \partial_t + \omega \,\partial_\phi$ with $\omega = \mathrm{const}$, the 'helical' Killing vector corresponding to rigid rotations at angular frequency $\omega$. Then $V^2 = 1 - \omega^2 \rho^2$ where $\rho^2 = (x^1)^2 + (x^2)^2$ and we have set $c = 1$. Furthermore
\begin{equation}
h_{ij} dx^i dx^j = \delta_{ij} dx^i dx^j + \frac{\omega^2}{1 - \omega^2 \rho^2}(x^1 dx^2 - x^2 dx^1)^2\,\,\,\,\,\mathrm{for}\,\,\rho < \frac{1}{\omega}\,.
\end{equation}
Here $D_i V$ describes the centrifugal force. One can \cite{beig2005relativistic} study (\ref{reduced}) for small $\Omega$ (for which $V=1$ and $h_{ij} = \delta_{ij}$) and $f$ close to a stressfree state $\mathring{f}$ (for which $t_i{}^j$ vanishes). Now the linearized operator given by the pair
\begin{equation}
\delta f^A \mapsto \left(\partial_j \delta_f t_i{}^j, \,\delta_f t_i{}^j \,n_j|_{\mathring{f}^{- 1}(\mathcal{\partial B})}\right)
\end{equation}
has a kernel consisting (at least) of the 6 elements of the form $\delta f^A = \xi^i \partial_i \mathring{f}^A$, where $\xi^i$ is a Euclidean Killing vector. This operator also does not have full range, but the set of pairs $(b_i, \tau_i)$ in its image, due to the symmetry of $t_{ij}$ plus the Killing equation, has to satisfy Eq. (\ref{corange}) namely
\begin{equation}\label{corange1}
\int_{\mathring{f}^{- 1}(\mathcal{B})} \xi^i b_i \,d^3\!x + \int_{\mathring{f}^{- 1}(\partial \mathcal{B})} \xi^i \tau_i \, dS(x) = 0\,,
\end{equation}
so the implicit function theorem does not directly apply. Still one can show existence of solutions for small $\omega^2$ (see \cite{beig2005relativistic}), provided that there hold
the constitutive conditions in Sect. 6 with pointwise stability on the elastic constants. Furthermore: the undeformed body in space, i.e. the domain $\mathring{f}^{- 1}(\mathcal{B})$ should be such that the rotation axis (a) goes through the centroid and (b) coincides with an axis of inertia. Finally (this can be relaxed) the 3 axes of inertia should be all different\footnote{The additional condition (97) in \cite{beig2005relativistic} turns out to be superfluous.}. Similar results \cite{beig2017celestial} can be proved for a sufficiently small body moving along a circular geodesic in Schwarzschild spacetime (note that such circular geodesics are orbits of a helical Killing vector).
\section{Coupling to gravity}
\subsection{Time independent case}
The problems in the previous section with constructing time independent solutions near stressfree configurations was ultimately due to the fact that these systems were not closed: an external force acted on the system, so precautions had to be made in order for this force to lie in the non-trivial range of the linearized operator. When gravity is coupled the system becomes closed, so one might expect a difference - and there is. For simplicity we explain this difference in the Newtonian case, see the end of Sect.6. Here the equations are (when $\mu(f) = 1$)
\begin{equation}\label{first}
\partial_j t_i{}^j = - n \,\partial_i U\,\,\,\mathrm{in}\,\,\,f^{- 1}(\mathcal{B}) \subset \mathbb{R}^3\,,\hspace{1cm}t_i{}^j n_j|_{f^{- 1}(\partial \mathcal{B})} = 0
\end{equation}
with $t_{ij} = 2 n \,\frac{\partial W}{\partial H^{AB}}\,(\partial_i f^A) (\partial_j f^B)$, $n = \mathrm{det}(\partial f)$ and
\begin{equation}\label{poisson}
\Delta U =
4 \pi G \,n\, \chi_{f^{- 1}(\mathcal{B})}\,,\hspace{1cm}U \rightarrow 0\,\,\mathrm{at}\,\infty
\end{equation}
The range problem just mentioned results from the fact that (\ref{first}) implies
\begin{equation}\label{2nd}
\int_{f^{- 1}(\mathcal{B})}\!\!n\,\xi^i \partial_i U\, d^3x = 0
\end{equation}
for the 6 Euclidean Killing vectors. Put differently, the total self-force and the total self-torque on the body due to gravity should be zero. But that is in fact true due to (\ref{poisson}), as can either be checked explicitly from
\begin{equation}
U(x) = - G \int_{f^{- 1}(\mathcal{B})} \frac{n(x')}{|x - x'|} d^3 x'
\end{equation}
or by first recalling
\begin{equation}
\partial_j \Theta^{ij} = n\, \partial^i U \,\chi_{f^{- 1}(\mathcal{B})}
\end{equation}
Now integrate the l.h. side against $\xi^i$ over $\mathbb{R}^3$: that is finite since $\xi = O(r), \Theta^{ij} = O(r^{- 4})$ and gives zero after partial integration and using the Killing equations. Using this fact one obtains \cite{beig2003static} an existence theorem for small $G$ and $f^A$ close to a stressfree configuration under appropriate conditions on the stored energy $W$. In spherical symmetry and for bodies $\mathcal{B}$ with the topology of a shell an analogous theorem was proved in \cite{kroon2006static}. Calogero and Leonori in \cite{calogero2015ground} obtained solutions without these smallness assumptions, using the calculus of variations.\\
The Einstein-elastic equations for an elastic body in an asymptotically flat gravitational field are a more complicated matter. But an argument in a similar spirit still works, where the Bianchi identity plays a crucial role. In this manner Andersson et al \cite{andersson2008static} obtained existence theorems for static elastic bodies with small $G$ and with the configuration close to stressfree, similarly for rigidly rotating ones for small $G, \omega$ in \cite{andersson2010rotating}. In the former case that provided the first construction of static self-gravitating bodies in GR without symmetries.\\
In spherical symmetry the subject started with \cite{magli1992generalization} (see also \cite{park2000spherically}, \cite{karlovini2003elastic} and \cite{frauendiener2007static}) and has meanwhile advanced much further, see the recent \cite{alho2022compact,alho2022erratum} and references therein. In these works there is no assumption of closeness to a stressfree configurations.
\subsection{Dynamics}
We finally make some remarks on the evolution problem for (possibly several) elastic bodies interacting through Einstein gravity. For more material consult the review paper \cite{andersson2015self}. The problem here lies mainly in the following facts.\\
The first problem is that of 'corner conditions' for the matter variable, i.e. $f^A$. Namely the matter boundary condition
\begin{equation}\label{mbound}
T_\mu{}^\nu m_\nu|_{f^{- 1}(\partial \mathcal{B})} = 0
\end{equation}
in order for a sufficiently regular solution to exist, requires that the initial data for $f^A$  satisfy the restriction to the initial surface and (\ref{mbound}) and that of a sufficient number of its time (i.e. tangential-to-$f^{- 1}(\partial \mathcal{B})$ derivatives): this is already present for bodies moving in a given background and is addressed in \cite{koch1993mixed,beig2007motion}\footnote{There is an incorrect argument in the Appendix of \cite{beig2007motion}, which can easily be remedied.}.\\
The second problem comes from what Andersson, Oliynyk call the transmission conditions \cite{andersson2014transmission}: these, loosely speaking result from the regularity condition mentioned earlier that first and second fundamental form of the metric behave continuously across the timelike surface $f^{- 1}(\partial \mathcal{B})$. In a similar vein as with the matter variable  this condition implies a sequence of - in this case - regularity conditions on the gravitational initial data. The corresponding evolution  problem has been solved in the works \cite{andersson2014transmission,andersson2016dynamical}, but compatibility with the GR initial value constraints is still an open issue.
\subsection*{Acknowledgement}
I am indebted to my long-term collaborator Bernd G. Schmidt for initiating my interest in this field 22 years ago.

\bibliographystyle{unsrt}

\end{document}